\documentclass[aps,prl,showpacs,amssymb,superscriptaddress, notitlepage,floats,floatfix,nofootinbib, twocolumn,groupedaddress]{revtex4-1}

\usepackage{graphicx}
\usepackage{dcolumn}
\usepackage{bm}
\usepackage{soul}
\usepackage{amsmath}
\usepackage{color}
\usepackage[dvipsnames]{xcolor}
\usepackage{hyperref}
\hypersetup{colorlinks=true, citecolor=MidnightBlue,
            linkcolor=CornflowerBlue, urlcolor=CornflowerBlue, linktocpage=true}

\newcommand{\ee}{\mathrm{e}}

\newcommand{\dd}{\mathrm{d}}

\newcommand{\Dl}{\nabla}

\def\apjl{Astrophys. J. Lett.}%
\def\apjs{Astrophys. J. Suppl.}%
\def\prd{Phys.~Rev.~D}%
\begin{document}

\title{EHT tests of the strong-field regime of General Relativity}
\author{Sebastian H. V\"olkel$^{1,2}$}
\email{svoelkel@sissa.it}
\author{Enrico Barausse$^{1,2}$}
\email{barausse@sissa.it}
\author{Nicola Franchini$^{1,2}$}
\email{nfranchi@sissa.it}
\author{Avery~E.~Broderick$^{3,4,5}$}
\email{abroderick@perimeterinstitute.ca}

\affiliation{$^1$SISSA, Via Bonomea 265, 34136 Trieste, Italy and INFN Sezione di Trieste\\
$^2$IFPU - Institute for Fundamental Physics of the Universe, Via Beirut 2, 34014 Trieste, Italy\\
$^3$Department of Physics and Astronomy, University of Waterloo, 200 University Avenue West, Waterloo, ON N2L 3G1, Canada\\
$^4$Perimeter Institute for Theoretical Physics, 31 Caroline Street North, Waterloo, ON N2L 2Y5, Canada\\
$^5$Waterloo Centre for Astrophysics, University of Waterloo, Waterloo, ON N2L 3G1, Canada}

\date{\today}

\begin{abstract}
Following up on a recent analysis by Psaltis {\it et al.} [Phys. Rev. Lett. 125, 141104 (2020)], we show 
that the observed shadow size of M87$^*$ can be used to unambiguously and robustly
constrain the black hole geometry in the vicinity of the circular photon orbit. Constraints
on the post-Newtonian weak-field expansion of the black hole's metric
are instead more subtle to obtain and interpret, as they rely on combining the shadow-size measurement
with suitable theoretical priors. We provide examples showing that post-Newtonian constraints 
resulting from shadow-size measurements should be handled with extreme care. 
We also discuss the similarities and complementarity between the EHT shadow
measurements and black-hole gravitational quasi-normal modes.
\end{abstract}

\maketitle

Until the LIGO/Virgo detection of gravitational waves (GWs)~\cite{PhysRevLett.116.061102,PhysRevLett.116.241103,PhysRevLett.118.221101,Abbott_2017,PhysRevLett.119.141101,Abbott_2020,LIGOScientific:2020stg,Abbott:2020khf,Abbott:2020tfl}, General Relativity (GR) had only been tested in
the  solar system (characterized by weak gravitational fields and mildly relativistic velocities $v$)~\cite{Will:2018bme} and in binary pulsars (which have strong gravitational fields inside/near the two stars,
but again mildly relativistic orbital velocities $v$)~\cite{1992PhRvD..45.1840D}. Since in both cases $v\ll c$,
 these tests  are normally performed within the post-Newtonian
formalism (i.e. an expansion in powers of $v/c$)~\cite{Blanchet:2013haa}.
The advent of GW astronomy has pushed
these tests to the strong-field \textit{and} highly relativistic regime that characterizes merging BH binaries~\cite{PhysRevLett.116.221101,LIGOScientific:2019fpa,Abbott:2020jks}, where 
the PN formalism breaks down
(except in the early-inspiral phase).

After coalescence, the  BH merger remnant
is expected to ``ring down'' by emitting quasi-normal modes (QNMs)~\cite{Kokkotas:1999bd,Nollert_1999,Berti_2009},
i.e. damped oscillations with discrete frequencies and
damping times,   functions of the BH mass and spin only (because of
the no-hair theorem~\cite{Israel:1967wq,Hawking:1971vc,Carter:1971zc,Robinson:1975bv}). 
Therefore, by measuring two QNMs, one can in 
principle test the no-hair theorem and thus GR~\cite{Dreyer:2003bv,Berti_2009}.
These tests are not yet possible with current detectors~\cite{Berti:2016lat} (even though 
there are clues of a second mode -- and namely the first overtone of the dominant mode -- besides the dominant QNM~\cite{Giesler:2019uxc}). However, comparatively weaker inspiral-ringdown self-consistency tests,
which compare the post-merger data to the signal predicted by GR by extrapolating the inspiral, 
show currently no hints of deviations from binary BHs in GR~\cite{PhysRevLett.116.221101,LIGOScientific:2019fpa,Abbott:2020jks}.

On larger scales, Psaltis {\it et al.}~\cite{PhysRevLett.125.141104}  proposed using the shadow size of M$87^*$ to constrain
deviations of the BH geometry from GR (i.e.~to test the no-hair theorem). The EHT shadow-size measurement is consistent
(to within 17\% at 68-percentile confidence level) with
the GR prediction\footnote{Like \cite{PhysRevLett.125.141104}, we assume that the shadow rim is associated with
 the photon sphere (i.e., for Schwarzschild BHs, the circular photon orbit). 
 This is supported by simulations  
  of the accretion flow onto  M$87^*$~\cite{Akiyama:2019fyp}. If this identification is not exact (see \S5.3, 7.3, and Appendix D of \cite{Akiyama:2019fyp}, cf. \cite{Gralla:2019xty,Gralla:2020pra}), it will lead to larger errors in our analysis and in that of \cite{PhysRevLett.125.141104}.
 } based on the object's mass-to-distance ratio derived from stellar dynamics~\cite{eht,Akiyama:2019eap,PhysRevLett.125.141104}.
Therefore, \cite{PhysRevLett.125.141104} concludes that theories extending/modifying GR cannot yield shadow sizes differing by more
than 17\% (at 68\% confidence) from the GR (i.e. Schwarzschild/Kerr) prediction~\cite{PhysRevLett.125.141104}.\footnote{This optimistically neglects
possible correlations of the parameters regulating deviations from GR with those of the BH (mass and spin) and accretion flow, which are  fixed in~\cite{PhysRevLett.125.141104}. These correlations may be important, as shown e.g. in \cite{Cardenas-Avendano:2019pec} for tests of GR with X-ray observations.}

While this idea is not new~\cite{Takahashi:2005hy,Johannsen:2010ru,Psaltis:2010ca,Amarilla:2011fx,Loeb:2013lfa,Psaltis:2014mca,Johannsen:2015hib,Psaltis:2015uza,Cunha:2015yba,Cunha:2016wzk,Psaltis:2018xkc,Cunha:2019dwb,Medeiros:2019cde}, Psaltis {\it et al.}~\cite{PhysRevLett.125.141104} used it to constrain  the BH geometry at 2PN order and beyond. 
To allow for deviations of the BH geometry from the Schwarzschild/Kerr solutions of GR, \cite{PhysRevLett.125.141104} utilizes parametrized metrics~\cite{Johannsen:2011dh,PhysRevD.87.124017,PhysRevD.83.104027,PhysRevD.88.044002,PhysRevD.90.084009} that yield the usual PN expansion far from the BH, but which might be valid also
in the  strong-field region, where the PN expansion fails. {\cite{PhysRevLett.125.141104} also assumes that {\it only one} of the parameters that regulate deviations from GR in these metrics is non-zero and allowed to vary.}

We will   show that  bounds on PN coefficients such as those obtained by \cite{PhysRevLett.125.141104}  depend sensitively on
the assumed form of the parametrized BH metric. This is not surprising,
as the results of \cite{PhysRevLett.125.141104} rely on a single measurement (the shadow's size).
We will show instead that EHT shadow-size measurements can robustly constrain
the BH geometry in the strong-field regime (near the circular photon orbit) and that these
constraints are complementary to those from GW observations of QNMs.

Throughout this paper, we use units $G=c=1$ and metric signature $-$$+$$+$$+$.

\textit{EHT constraints on the PN expansion of BH geometries:}
As an example, we focus on the non-spinning parametrized geometry of Rezzolla and Zhidenko (RZ) \cite{PhysRevD.90.084009} (see  \cite{Konoplya:2016jvv,Younsi:2016azx} for the axisymmetric generalization).\footnote{We neglect the spin as our goal is to discuss the robustness of
the M87$^*$ shadow-size constraints with respect to the parametrization of the non-GR effects.}
The  $g_{tt}$ component  is 
\begin{align}
	-g_{tt} =&  x \Big(1 - \varepsilon(1-x) + (a_0 -\varepsilon)(1-x)^2 \nonumber \\ &+ \tilde{A}(x)(1-x)^3 \Big),\label{RZ_metric}
\end{align}
with $ x \equiv 1- r_0/r$ ($r_0$ being the horizon's radius),  and 
\begin{align}\label{A_cf}
	\varepsilon = - \left(1-\frac{2M}{r_0} \right), \qquad
	\tilde{A}(x) = \cfrac{a_1}{1+\cfrac{a_2x}{1+\cfrac{a_3x}{1+\dots}}}\,,
\end{align}
where $M$ is the mass and the dots represent a continued fraction structure. The Schwarzschild
limit is recovered when $\varepsilon,\,a_i\to0$ (with $i=0,1,2,3,\dots$). When expanded in 
orders of $1/r$, one recovers the usual PN structure
\begin{align}\label{PNgtt}
	-g_{tt}  = 1 - \frac{2M}{r} + \sum_{i=1}^{\infty}P_i\left(\frac{M}{r}\right)^{i+1},
\end{align}
where the parameters $P_i$ depend on  $\varepsilon,\,a_i$ and vanish in the Schwarzschild limit. However,
the continued-fraction structure of Eq.~\eqref{A_cf} is introduced to accelerate the convergence of the PN expansion (i.e. to ``resum'' it),
so that the RZ metric aims to provide an accurate description even in the strong-field regime.

Null geodesics  in a static, spherically symmetric geometry satisfy \cite{Bardeen:1973tla}
\begin{align}\label{geodesic_potential}
&\frac{(g_{tt})^2}{E^2} \left(\frac{\text{d}r_*}{\text{d} \sigma} \right)^2+V=0\,,\\
&V=-1-\frac{b^2 g_{tt}}{r^2}\,,
\end{align}
where ${\text{d}r_*}/{\text{d} \sigma}$ is the
derivative (with respect to an affine parameter $\sigma$) of the tortoise coordinate $r_*$ (defined by $\dd r_* / \dd r = \sqrt{-g_{rr}/g_{tt}}$, with $r$ the areal radius), and $b=L/E$ (with $E$ and $L$  the photon's conserved energy and angular
momentum) is the impact parameter. The circular photon orbit's radius, $r_\text{ph}$, and its impact parameter, $b_{\rm ph}$,
correspond to a minimum of the effective potential, i.e. they solve 
$V={\rm d}V/{\rm d} r=0$.  These conditions can be  reduced to~\cite{PhysRevLett.125.141104}
\begin{align}\label{f_ph}
r_{\rm ph}&=\left(\frac{\text{d}}{\text{d}r} \ln \sqrt{-g_{tt}}\right)^{-1} \Bigg|_{r_{\rm ph}},\\
	b_\text{ph}  &= \frac{r_\text{ph}}{\sqrt{-g_{tt}(r_\text{ph})}},\label{shadow}
\end{align}
which can be solved numerically for generic metrics; the Schwarzschild metric gives $r_{\rm ph}=3 M$ and $b_{\rm ph}=3 \sqrt{3} M$. If $b_{\rm ph}$ is 
interpreted as the measured shadow size (like in \cite{PhysRevLett.125.141104}), the EHT shadow-size measurement bounds $|b_{\rm ph}/M-3 \sqrt{3}|/(3 \sqrt{3})\lesssim 0.17$ at 68\% confidence~\cite{PhysRevLett.125.141104}.

\begin{figure}
	\centering
	\includegraphics[width=1.0\linewidth]{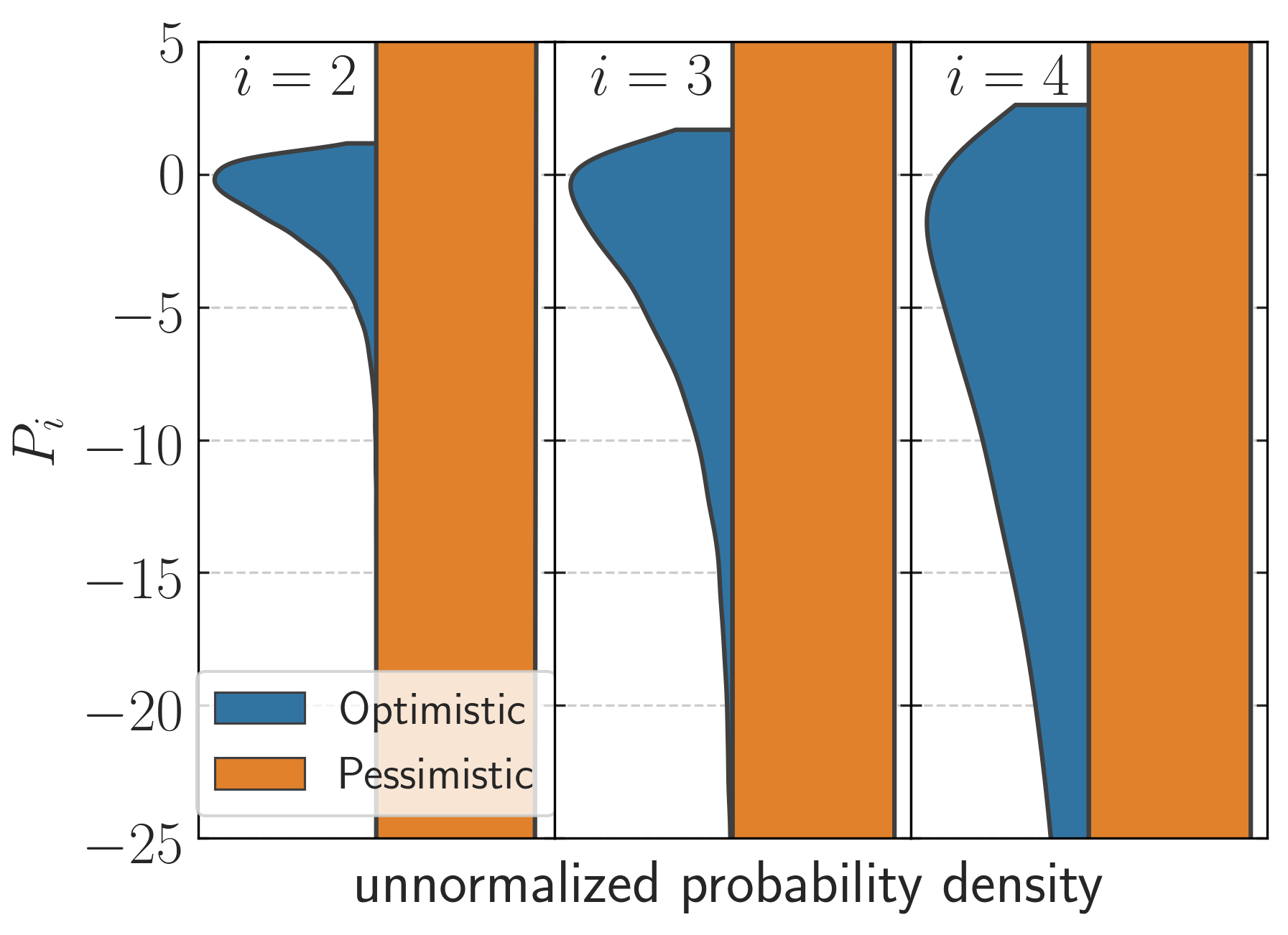}
	\caption{Posteriors of the 2PN  (left violin plot), 3PN (middle) and 4PN coefficients of the BH metric. The left half of each plot shows the ``optimistic'' bound obtained when only the term under consideration is allowed to deviate from GR, while the right half shows the posterior distribution for the same quantity, but in a parametrized metric with four parameters $\epsilon$, $a_1$, $a_2$ and $a_3$. The sharp cutoff of the ``optimistic'' posteriors is \textit{not} an artifact, and roughly corresponds to the metric ceasing to possess an event horizon. The ``pessimistic'' posteriors coincide with the priors.}
	 \label{violins}
\end{figure}

The implications of this bound for the BH geometry depend on our prior knowledge of its functional form. For instance, if we assume Eq.~\eqref{PNgtt} with
all PN coefficients set to zero except for the 2PN term, i.e. $P_i=0$ for $i\neq 2$
and flat priors for $P_2$, one
obtains the posterior distribution for $P_2$ shown in the first panel (left half) of Fig.~\ref{violins}. However, consider the RZ metric, 
imposing agreement with GR at 1PN order ($a_0=0$; we will comment on
this assumption below) and setting $a_i=0$ for $i\geq4$ for simplicity. With the four non-zero parameters $\epsilon$, $a_1$, $a_2$ and $a_3$ (for which we assume large flat priors of $\pm 30$),  the posterior distribution
for 
\begin{equation}
P_2= -\frac{r_0^2}{M^2} \left[\left(-\frac{a_1 \left(a_3+1\right)}{a_2+a_3+1}+1\right) \frac{r_0}{M}-2 \right]
\end{equation}
is extremely broad (first panel of Fig.~\ref{violins}, right half). Similar conclusions hold for the higher PN terms whose expressions we do not show as lengthy and uninformative. In the second and third panels of Fig.~\ref{violins}, we show similar bounds for the 3PN and 4PN coefficients $P_3$ and $P_4$, both when they are considered optimistically as the only free parameters [one at a time, via the metric \eqref{PNgtt}], and when their posteriors are instead obtained from those of the parameters $\epsilon$, $a_1$, $a_2$ and $a_3$. All  results of Fig.~\ref{violins} were produced with the   Metropolis-Hastings sampler of PyMC3~\cite{pymc3}, assuming Gaussian errors on the EHT shadow-size measurement. 
The bounds of Fig.~\ref{violins} are reminiscent of those presented in Fig. 7 of \cite{PhysRevLett.116.221101} by the LIGO/Virgo collaboration for the PN coefficients of the BH-binary inspiral GW signal.\footnote{See also \cite{xray19_12} for  GW bounds on
the PN coefficients of the conservative dynamics, which
are stronger than  those claimed by \cite{PhysRevLett.125.141104} although comparable directly to them (because like \cite{PhysRevLett.125.141104}, \cite{xray19_12} assumes only one free
non-GR parameter).} Indeed, 
our approach resembles closely that of \cite{PhysRevLett.116.221101}, i.e. ``optimistic'' bounds are obtained by letting those PN coefficients free one by one, while ``pessimistic'' bounds assume that they are all allowed to vary simultaneously. We stress that correlations between the PN coefficients,
when several of them are allowed to vary simultaneously, also
appear in the case of GW measurements. 
However,
 unlike our case, the LIGO/Virgo ``pessimistic''  posteriors are smaller than the priors, at least for the leading-order (0PN) term in the GW phase, cf. Table I of \cite{PhysRevLett.116.221101}
 (see also \cite{Sampson:2013lpa,Yunes:2016jcc,Psaltis:2020ctj}).
 This will be even more true for the -1PN term~\cite{Barausse:2016eii,LIGOScientific:2019fpa}.
  Note 
 also that the conclusions of our Fig.~\ref{violins} (and namely that the ``pessimistic'' bounds on the single PN parameters---while marginalizing on all others---coincide with the priors) are robust against inclusion of lower-PN orders (e.g. $P_1$).

 Fig.~\ref{violins} shows that the bounds of \cite{PhysRevLett.125.141104} are not robust, but  depend on the  form of the parametrized metric and on the priors on its parameters. To illustrate how subtle it is to put priors on the shape  and PN coefficients of parametrized metrics, let us stress that even
 though we have followed above
 \cite{PhysRevLett.125.141104} and  required $g_{tt}$
 to match the Schwarzschild solution at 1PN order (i.e. $a_0=P_1=0$), there is in principle no reason to do so.
  While \cite{PhysRevLett.125.141104} sets
 1PN deviations from GR to zero, theories different from GR do not necessarily satisfy  Birkhoff's theorem~\cite{Berti_review} and may not obey the strong-equivalence principle~\cite{Barausse:2016eii}. Therefore,
 the metric around BHs need not be the same as around a star, and one
 cannot invoke solar-system tests to set the 1PN deviations from GR to zero. While this was mentioned in  \cite{PhysRevLett.125.141104}, the implications
 of the assumption of vanishing 1PN deviations from GR was
 not explored (\cite{PhysRevLett.125.141104} simply
 mentions that as a ``very conservative'' assumption). We will now explore cases where that assumption is not verified and affects the bounds that one obtains. 
 
 An  example is given by a theory with a ``dark photon'' [i.e. a U(1) gauge field], possibly coupled with a scalar~\cite{Garfinkle:1990qj,Hirschmann:2017psw}:
 \begin{equation}
 S= \int \frac{\dd^4 x \sqrt{-g}}{16\pi} \left[R - \frac{1}{2}\Dl_a\phi\Dl^a\phi + \ee^{-\alpha_0\phi}F_{ab}F^{ab} \right] 
 \end{equation}
  This is known as Einstein-Maxwell-dilaton theory [if the U(1) symmetry is broken, e.g. by a light mass, the dark photon may even be the dark matter; cf.~\cite{McDermott:2019lch} and references therein]. Spherical BHs in this theory are described by a generalized Reissner-Nordstr{\"o}m metric. The $tt$ component reads~\cite{Gibbons:1987ps,Garfinkle:1990qj} 
   \begin{equation}\label{gttRN}
      -g_{tt} = \left( 1 - \frac{r_+}{\bar{r}} \right) \left( 1 - \frac{r_-}{\bar{r}} \right)^{1-\alpha_1}\,,
   \end{equation}
where the areal radius is  $r = \bar{r} (1 - r_-/\bar{r})^{\alpha_1/2}$, $\alpha_1 = 2\alpha_0^2/(1+\alpha_0^2)$ and the constants $r_\pm$ are related to the mass $M$ and the dark charge $Q$ of the BH through $2M  = r_+ + (1-\alpha_1)r_-$ and $2Q^2  = r_+ r_- (2-\alpha_1)$.
 Unlike  an electric charge, which is  neutralized by the plasma near the horizon~\cite{env_effects}, $Q$ may be non-zero
 and different for a BH (where it is a free parameter)
 and a star (where it vanishes unless the dark photon and/or the scalar are 
 coupled to the Standard Model). Obviously, for $Q\neq0$ and $\alpha_1 \neq 1$ the metric \eqref{gttRN} deviates from Schwarzschild at 1PN order.

\begin{figure}
	\centering
	\includegraphics[width=1.0\linewidth]{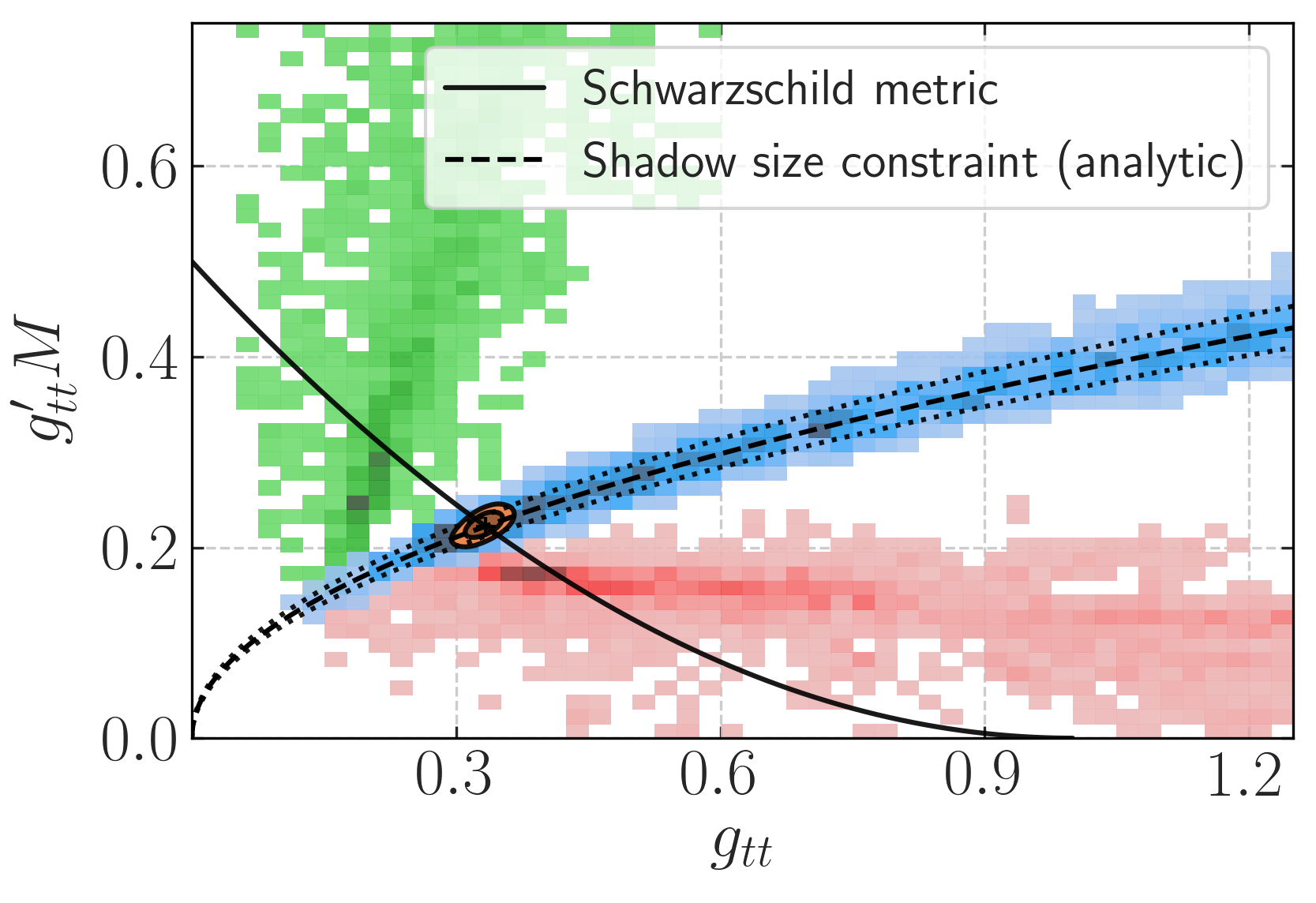}
	\caption{Posteriors for $g_{tt}$ and $g^\prime_{tt}$,
	produced by our MCMC sampler in the case of shadow-size measurement with 5\% error,
	and five free parameters $\epsilon, a_0, a_1, a_2, a_3$. The posteriors are
	evaluated at $r=r_\mathrm{ph}$ (blue), $r=0.85\,r_\mathrm{ph}$ (green) and $r=1.15\,r_\mathrm{ph}$ (red). Analytic predictions for $g_{tt}$ and $g^\prime_{tt}$
	at $r_{\rm ph}$ [from Eqs.~\eqref{f_ph} and~\eqref{shadow}] are provided by the black dashed line and by the black dotted ones ($1\sigma$ error bars). 
	For comparison, 
the Schwarzschild metric is shown for $r\in [2M, \infty]$ (black solid). To demonstrate the close relation between QNM and  shadow-size measurements, we also show the MCMC results 
assuming a $l=2, n=0$ QNM measured with $5\%$ accuracy (orange), using the code of  Ref.~\cite{paper10}.}
	 \label{gttvsdgtt}
\end{figure}

\textit{EHT strong-field tests of gravity:}
We stress that the behavior 
of Fig.~\ref{violins} -- and namely the fact that the (marginalized)
pessimistic bounds on the PN coefficients coincide with the priors --
is not only due to the larger number of parameters being varied.
In fact, irrespective of how many parameters the model (i.e. the parametrized metric) has,
the shadow size does constrain a particular combination of the parameters. We show this explicitly in Fig.~\ref{gttvsdgtt}, where we plot the posteriors for $g_{tt}$
and ${\rm d} g_{tt}/{\rm d} r$ (evaluated at  $r_{\rm ph}$) in the ``pessimistic'' scenario of Fig.~\ref{violins}.\footnote{We assume here a 5\% precision for the shadow-size measurement, which may be achievable with next-generation EHT-like experiments~\cite{2021ApJS..253....5R}, to show that these conclusions will not change with better future data.}
This 
shows that the data \textit{is} informative and that the inference can be robust against the number of parameters involved, as long as one asks the right question (i.e. as long as one does not attempt to estimate the PN coefficients, but focuses instead on the geometry near the circular photon orbit).
Conversely, the geometry away from the circular photon orbit (even in its immediate vicinity)
is less constrained, as shown by the posteriors
at $r=0.85 r_{\rm ph}$ and $r=1.15 r_{\rm ph}$ plotted in Fig.~\ref{gttvsdgtt}. This plot therefore highlights that the 
behavior of the PN constraints 
of Fig.~\ref{violins} is due to the slow
convergence (if any) of 
 PN expansion near the circular photon orbit.

As correctly stated in \cite{PhysRevLett.125.141104},   ``if more than one PN parameter
[...] is included, then the size
measurement of the BH shadow will [...] lead
to a constraint on a linear combination of these parameters.''
However, the  combination in question is nothing but $b_{\rm ph}$ itself (as shown by the solid and dashed lines in  Fig.~\ref{gttvsdgtt}), and becomes 
linear only if deviations of the PN coefficients from their GR values are small (which is not obvious). Under this assumption, however, one can solve Eqs.~\eqref{f_ph}--\eqref{shadow}
by inserting the PN metric \eqref{PNgtt}
and linearizing in the $P_i$. One then obtains that the EHT bound
is approximately (at 68\% confidence and assuming vanishing BH spin)
\begin{equation}\label{combo}
\frac12 \left\vert\sum_{i=1}^{\infty}\frac{P_i}{3^i}\right\vert\lesssim 0.17\,.
\end{equation}
We stress that the coefficients of this combination will be different for non-vanishing spin. In fact, since for high spins
the circular photon orbit approaches (in Boyer-Lindquist coordinates) the horizon~\cite{Bardeen:1973tla}, we expect those coefficients to all be comparable.

This approximate constraint explains the growing width of
the ``optimistic'' bounds of Fig.~\ref{violins} as the PN order increases. It also suggests that bounds on the lowest-order PN parameters (0PN~\cite{PhysRevLett.116.221101} and -1PN~\cite{Barausse:2016eii,LIGOScientific:2019fpa})
from GW observations of the early inspiral of BH binaries are 
potentially stronger than than those from shadow-size measurements, even though posterior correlations may still appear at higher-PN orders, when several parameters are varied at the same time~\cite{PhysRevLett.116.221101,Shoom:2021mdj}. In more detail, GW detectors are sensitive
to a whole time (or frequency) series, unlike the EHT shadow-size observation (which amounts to a single data point). They measure, in particular, 
 the GW phase 
$\Phi(f)=\Phi_{\rm GR}(f) [\sum^{\infty}_{i=-1}\delta_i (M f)^{2i/3}]$, where $\Phi_{\rm GR}(f) $ is the GR-predicted phase, $f$
 the GW frequency, $M$ the binary total mass, and the $\delta_i$ are parameters accounting for deviations from GR at (integer) PN orders~\cite{ppe1,ppe2,pp3,Yunes:2016jcc,PhysRevLett.116.221101,LIGOScientific:2019fpa,xray19_12} (in the GR limit, $\delta_0=1$ and $\delta_i=0$ for $i>0$).
Since $(Mf)^{2/3}\propto M/r$ (with $r$ the orbital separation), the phase is a series in $M/r$, just like Eq.~\eqref{combo}. The difference with 
Eq.~\eqref{combo} is that BH-binary inspiral observations are sensitive to separations from tens of $M$ down to $\sim 6M$, which accelerates convergence.
Similarly,  X-ray observations of accretion disks around BHs~\cite{Bambi:2011jq,Bambi:2015kza,xray19_3,Nampalliwar:2019iti,xray19_12} are sensitive to radii larger than the innermost stable circular orbit, thus being in a regime where the PN expansion is applicable (although  bounds derived from these observations may
depend on  the accretion-disk model~\cite{extra1,extra2,Cardenas-Avendano:2019pec}).

\begin{figure}
	\centering
	\includegraphics[width=1.0\linewidth]{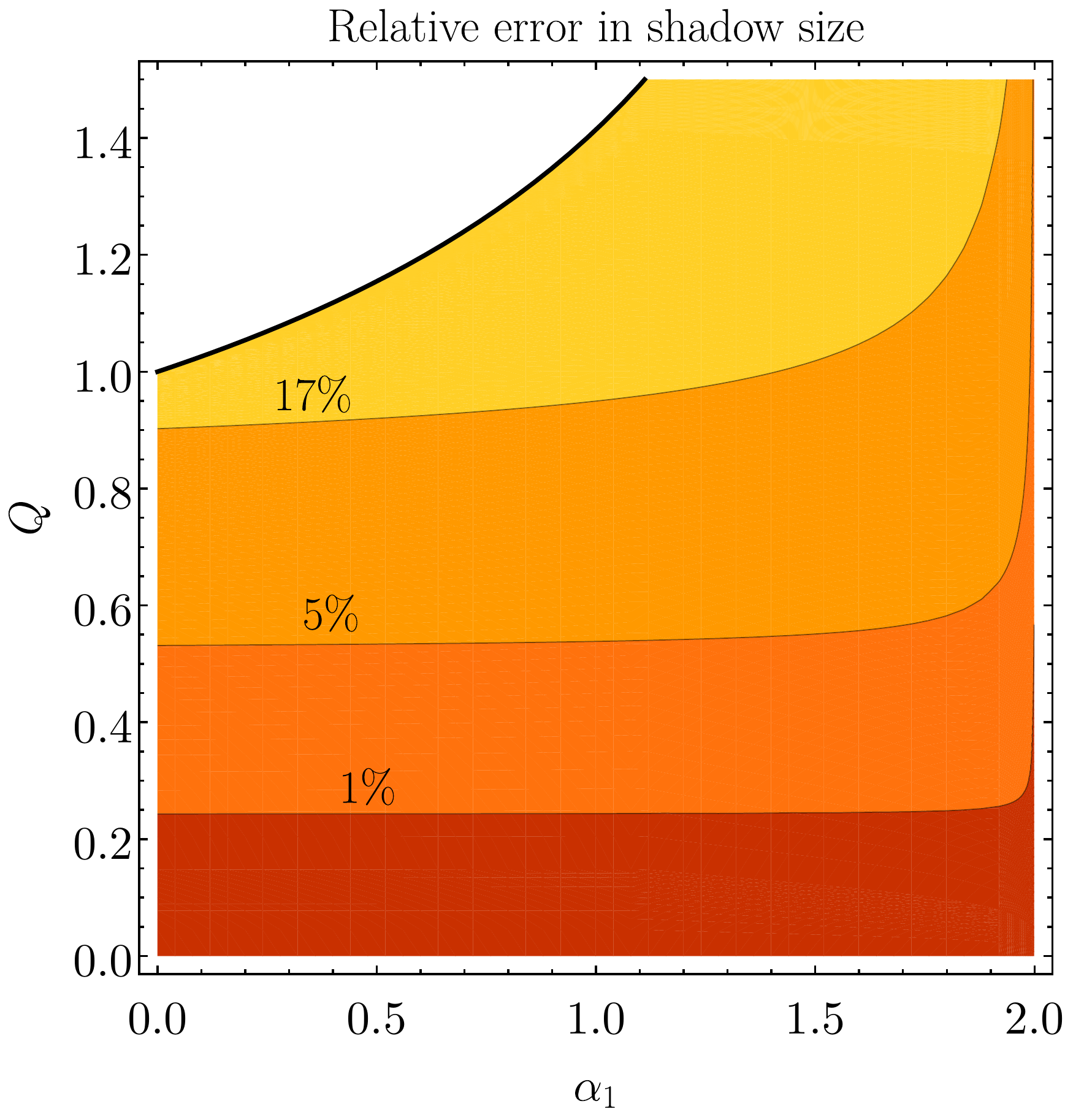}
	\caption{Relative difference between the
	shadow size of a BH in Einstein-Maxwell-dilaton gravity and a Schwarzschild BH with the same mass. Contours correspond to differences of $1\%$, $5\%$ and $17\%$ (from darker to lighter regions). The solid black line represents extremal solutions.  Reissner-Nordstr{\"o}m solutions lie on the line $\alpha_1=0$.
	 \label{shadowcontours}}
\end{figure}

An alternative way to  exploit  EHT observations  is to consider BH solutions in gravitational theories different from GR, which are often known exactly and/or numerically and which will in general show deviations from the Schwarzschild/Kerr metric at all PN orders. To show this explicitly, we will present  a few examples, making (like above) the simplifying assumption of spherical symmetry.\footnote{Such a theory-by-theory approach is not needed in situations where the PN expansion
is well suited for the system at hand and yields robust bounds. This is the case
e.g. for solar-system tests, for which parametric bounds on the 1PN coefficients
are robust and can be readily converted into constraints on specific theories.}
 Consider first the Reissner-Nordstr{\"o}m-like  BH of Eq.~\eqref{gttRN}.
 If $\alpha_1=0$, that reduces exactly to the Reissner-Nordstr{\"o}m spacetime, which features $P_i=0$ for $i\geq2$. The EHT shadow-size measurement then bounds $P_1\propto Q^2$  in the 68\% confidence interval $[0,0.81]$. In the general case, the solution is governed by two independent parameters: the coupling $\alpha_1$ and the BH charge $Q$. 
 The fractional difference between $b_{\rm ph}$ and the Schwarzschild
 value is shown in Fig.~\ref{shadowcontours}, with the
 region within the 17\% contour being in agreement with the 
  EHT shadow-size measurement. From these bounds, one may then obtain posteriors for the PN coefficients, whose  $68\%$ confidence intervals are $P_1\in[-1.04,0.81]$, $P_2\in[-0.90,0.12]$, $P_3\in[-1.40,0.053]$, $P_4\in[-2.48,0.031]$.

Hairy BHs differing from the Schwarzschild one can also be obtained in scalar-tensor theories, provided that a coupling between the scalar $\phi$ and the Gauss-Bonnet invariant $\mathcal{G} = R^{abcd}R_{abcd} - 4 R^{ab}R_{ab} + R^2$ is introduced, giving rise theories known as Einstein-scalar-Gauss-Bonnet gravity.
Their action is~\cite{Julie:2019sab}
\begin{equation}
	S = \int \frac{\dd^4 x \sqrt{-g}}{16\pi} \left[R - \frac12\Dl_a\phi\Dl^a\phi + \lambda^2 f(\phi) \mathcal{G} \right]\,,
\end{equation}
with $\lambda$ a coupling constant (with dimensions of a length) and $f(\phi)$ a dimensionless coupling function. Provided that
$f(\phi)$ is monotonic,
one can find BH solutions characterized by a dimensionless scalar charge, defined from the decay $\phi \simeq \phi_\infty + q M/r $ of the scalar near spatial infinity, being $M$ the BH mass and $\phi_\infty$ the asymptotic value of the scalar field. For these theories $q=\lambda^2 f'(\phi_\infty)/(2M^2)$, where a prime denotes the derivative with respect to $\phi$.
Numerical solutions can be found for specific coupling functions, e.g. $f(\phi) = \exp{(\phi)}$~\cite{Kanti:1995vq,Pani:2009wy}, or $f(\phi) \propto{\phi}$~\cite{Sotiriou:2013qea,Sotiriou:2014pfa}.
Alternatively, one can find solutions perturbatively in the charge $q$ for various monotonic coupling functions~\cite{Mignemi:1992nt,Yunes:2011we,Maselli:2015tta,Julie:2019sab}.
Using the fits of \cite{Kokkotas:2017ymc} to the numerical solutions of~\cite{Kanti:1995vq} [for $f(\phi) = \exp{(\phi)}$], we find that these BHs
always agree with the EHT shadow-size measurement if the precision of the measurement is worse than $\sim 1\%$. This confirms the findings of~\cite{Cunha:2016wzk}.

Interesting solutions can also be found  
for non-monotonic coupling functions. If
$f(\phi)$ has a minimum, BHs can scalarize spontaneously in Einstein-scalar-Gauss-Bonnet gravity~\cite{Silva:2017uqg,Doneva:2017bvd,Antoniou:2017acq,Cunha:2019dwb,Collodel:2019kkx,Herdeiro:2020wei,Berti:2020kgk}.
These scalarized BHs form because the Schwarzschild and/or Kerr solutions of GR become tachyonically unstable in these theories~\cite{Silva:2017uqg,Doneva:2017bvd,Andreou:2019ikc,Minamitsuji:2019iwp,Dima:2020yac,Doneva:2020nbb,Doneva:2020kfv}. 

Consider the case $f(\phi) = \left[1-\exp(-6\phi^2)\right]/12$, studied in \cite{Cunha:2019dwb,Doneva:2017bvd}. This theory provides a continuum set of scalarized BHs with mass $M \in [0,0.587 \lambda]$. The EHT shadow-size measurement  then constrains $\lambda < 3.27M$ (see also \cite{Cunha:2019dwb})\footnote{We expect $\lambda\gg 100 M_\odot$
to agree with early-inspiral GW observations (because 
$q\to0$ as $M/\lambda\sim0$: cf. 
\cite{Doneva:2017bvd}, Fig. 4), although merger/ringdown bounds  have not been derived yet.}. While this  bound  is not new (having been derived  in~\cite{Cunha:2019dwb}, cf. their Fig.~5), one can compare it with the results of our Fig.~\ref{violins} for the RZ parametrized metric. The PN expansion of scalarized BHs (obtained by solving the field equations perturbatively near spatial infinity) yields PN parameters $P_1=0$, $P_2 = q^2/3$, $P_3 = 2q^2 / 3$, $P_4 = 6q^2/5 - 3q^4/20 + 16 q^2 \lambda^2/5 M^2$. Moreover, an additional coupling of the scalar to the Ricci curvature can even give rise to $P_1\neq0$~\cite{Antoniou:2021zoy}. The charge $q$ is a function of $\lambda$, and can be extracted from the numerical solutions to the full field equations. The EHT bound  $\lambda/M < 3.27$ then translates into the constraint $ q <0.195 $ and thus $|P_2|<0.136$, $|P_3|<0.272$, $|P_4|<14.46$. As can be seen, in this specific case the bounds on the PN parameters 
are comparable to (or even better than) the ``optimistic'' bounds
 of  Fig.~\ref{violins}.

\textit{Discussion:}
We stress  that the dependence of the shadow-size bounds 
on the PN coefficients on how many of them 
 are allowed to vary, as well as the robustness of the constraints on the
 geometry near the circular photon orbit
are reminiscent of what happens with GW observations
of QNMs in the ringdown phase of binary BHs.
 Indeed, shadow-size measurements are to the EHT what QNMs are to GW detectors. Both shadows and QNMs are sensitive to the BH geometry near the circular photon orbit, and their physics cannot be described within the PN approximation. Note that  \cite{paper10} attempted to constrain parametrized metrics (e.g. the RZ metric) with QNM observations. In agreement with this letter, \cite{paper10} found that 
 constraints on the reconstructed geometry are  robust near the peak of the effective potential.
We show this explictly in Fig.~\ref{gttvsdgtt},
where we present projected QNM constraints on
$g_{tt}$ and ${\rm d} g_{tt}/{\rm d}r$, alongside those
from the shadow size mentioned earlier, in the ``pessimistic''
case where the four non-zero parameters $\epsilon$, $a_1$, $a_2$ and $a_3$
are allowed to vary simultaneously.

In more detail, the  geometric-optics limit
 of the GW propagation equation reduces, in GR, to the null-geodesics one 
(see e.g., \cite{paper10}), i.e., high-frequency
gravitational wavefronts follow null geodesics. Therefore, the effective potential
for QNMs in the  geometric-optics limit (i.e., the limit of large angular eigen-numbers $\ell,m$) coincides with that of null geodesics [Eq.~\eqref{geodesic_potential}]. Since QNMs are generated at the peak of the effective potential, which is close to the circular photon orbit (and coincides with it for $\ell\gtrsim m\gg 1$), it is not surprising that the QNM frequencies of the Kerr spacetime 
are given (in the geometric-optics limit) by linear combinations of
the orbital and frame-dragging precession
frequencies of the circular null orbit (or simply by multiples
of the orbital frequency in Schwarzschild, where the two frequencies coincide)~\cite{1972ApJ...172L..95G,wkb1,wkb2,Yang_2012}. Similarly, one can relate QNM decay times to the Lyapunov exponents of null
geodesics near the circular photon orbit~\cite{Cardoso:2008bp,wkb1,wkb2,Yang_2012}. These exponents
depend in turn on the curvature of the effective potential for photon orbits  near its peak.

This null
geodesics/QNMs correspondence
can be generalized to BH spacetimes different from Kerr/Schwarzschild, at least in a wide class of gravitational theories \cite{Glampedakis_2017,Glampedakis:2019dqh,Silva:2019scu,paper10}. This correspondence motivates combining 
  EHT shadow-size tests of the no-hair theorem with the QNM null tests of the same theorem that will become possible 
with third-generation GW interferometers or spaced-based detectors such as e.g., LISA~\cite{Berti:2016lat} or TianQin~\cite{Shi:2019hqa}.

\begin{acknowledgments}
\textit{Acknowledgments:} S.V., E.B. and N.F. acknowledge financial support provided under the European Union's H2020 ERC Consolidator Grant ``GRavity from Astrophysical to Microscopic Scales'' grant agreement no. GRAMS-815673. 
This work was supported in part by Perimeter Institute for Theoretical Physics.  Research at Perimeter Institute is supported by the Government of Canada through the Department of Innovation, Science and Economic Development Canada and by the Province of Ontario through the Ministry of Economic Development, Job Creation and Trade.
A.E.B. thanks the Delaney Family for their generous financial support via the Delaney Family John A. Wheeler Chair at Perimeter Institute.
A.E.B. and receives additional financial support from the Natural Sciences and Engineering Research Council of Canada through a Discovery Grant.
We thank A. Cardenas Avendano, E. Berti, K. Glampedakis, R. Gold, P. Kocherlakota, K. D. Kokkotas, L. Rezzolla, N. Wex and N. Yunes  for insightful conversations and for reviewing a draft of this manuscript. 
During the completion of this work we have become
aware of a related work by P. Kocherlakota, L. Rezzolla, {\it et al.}, which deals with topics that partly overlap with those of this manuscript (i.e. EHT bounds on exact non-GR BH solutions).
\end{acknowledgments}

\bibliography{literature}

\end{document}